\documentclass[12pt]{article}
\usepackage{fancyhdr}

\usepackage{mathrsfs}
\usepackage[T1]{fontenc}
\usepackage{mathpazo}
\usepackage{setspace}
\usepackage{amsfonts}
\usepackage{amssymb}
\usepackage{amsmath}
\usepackage{epsfig}
\usepackage{latexsym}
\usepackage{color}
\usepackage{graphicx}
\usepackage{nicefrac}
\usepackage[latin1]{inputenc}
\usepackage{slashed}
\usepackage{multirow}
\usepackage{comment}
\usepackage{soul}
\usepackage{hyperref}
\usepackage{cite}

\usepackage[titletoc]{appendix}

\def\hybrid{\topmargin -20pt    \oddsidemargin 0pt
        \headheight 0pt \headsep 0pt
        \textwidth 6.25in       
        \textheight 9.25in       
        \marginparwidth .875in
        \parskip 5pt plus 1pt   \jot = 1.5ex}

\hybrid

\def\baselinestretch{1.2}

\catcode`\@=11

\def\marginnote#1{}
\newcount\hour
\newcount\minute
\newtoks\amorpm
\hour=\time\divide\hour by60
\minute=\time{\multiply\hour by60 \global\advance\minute by-\hour}
\edef\standardtime{{\ifnum\hour<12 \global\amorpm={am}%
        \else\global\amorpm={pm}\advance\hour by-12 \fi
        \ifnum\hour=0 \hour=12 \fi
        \number\hour:\ifnum\minute<10 0\fi\number\minute\the\amorpm}}
\edef\militarytime{\number\hour:\ifnum\minute<10 0\fi\number\minute}

\def\draftlabel#1{{\@bsphack\if@filesw {\let\thepage\relax
   \xdef\@gtempa{\write\@auxout{\string
      \newlabel{#1}{{\@currentlabel}{\thepage}}}}}\@gtempa
   \if@nobreak \ifvmode\nobreak\fi\fi\fi\@esphack}
        \gdef\@eqnlabel{#1}}
\def\@eqnlabel{}
\def\@vacuum{}
\def\draftmarginnote#1{\marginpar{\raggedright\scriptsize\tt#1}}

\def\draft{\oddsidemargin -.5truein
        \def\@oddfoot{\sl preliminary draft \hfil
        \rm\thepage\hfil\sl\today\quad\militarytime}
        \let\@evenfoot\@oddfoot \overfullrule 3pt
        \let\label=\draftlabel
        \let\marginnote=\draftmarginnote
   \def\@eqnnum{(\theequation)\rlap{\kern\marginparsep\tt\@eqnlabel}%
\global\let\@eqnlabel\@vacuum}  }

\def\preprint{\twocolumn\sloppy\flushbottom\parindent 2em
        \leftmargini 2em\leftmarginv .5em\leftmarginvi .5em
        \oddsidemargin -.5in    \evensidemargin -.5in
        \columnsep .4in \footheight 0pt
        \textwidth 10.in        \topmargin  -.4in
        \headheight 12pt \topskip .4in
        \textheight 6.9in \footskip 0pt
        \def\@oddhead{\thepage\hfil\addtocounter{page}{1}\thepage}
        \let\@evenhead\@oddhead \def\@oddfoot{} \def\@evenfoot{} }

\def\numberbysection{\@addtoreset{equation}{section}
        \def\theequation{\thesection.\arabic{equation}}}

\def\underline#1{\relax\ifmmode\@@underline#1\else
        $\@@underline{\hbox{#1}}$\relax\fi}

\def\titlepage{\@restonecolfalse\if@twocolumn\@restonecoltrue\onecolumn
     \else \newpage \fi \thispagestyle{empty}\c@page\z@
        \def\thefootnote{\fnsymbol{footnote}} }

\def\endtitlepage{\if@restonecol\twocolumn \else \newpage \fi
        \def\thefootnote{\arabic{footnote}}
        \setcounter{footnote}{0}}  

\catcode`@=12
\relax

\def\figcap{\section*{Figure Captions\markboth
        {FIGURECAPTIONS}{FIGURECAPTIONS}}\list
        {Figure \arabic{enumi}:\hfill}{\settowidth\labelwidth{Figure
999:}
        \leftmargin\labelwidth
        \advance\leftmargin\labelsep\usecounter{enumi}}}
 \relax
\def\tablecap{\section*{Table Captions\markboth
        {TABLECAPTIONS}{TABLECAPTIONS}}\list
        {Table \arabic{enumi}:\hfill}{\settowidth\labelwidth{Table
999:}
        \leftmargin\labelwidth
        \advance\leftmargin\labelsep\usecounter{enumi}}}
 \relax
\def\reflist{\section*{References\markboth
        {REFLIST}{REFLIST}}\list
        {[\arabic{enumi}]\hfill}{\settowidth\labelwidth{[999]}
        \leftmargin\labelwidth
        \advance\leftmargin\labelsep\usecounter{enumi}}}
 \relax
\makeatletter
\newcounter{pubctr}
\def\publist{\@ifnextchar[{\@publist}{\@@publist}}
\def\@publist[#1]{\list
        {[\arabic{pubctr}]\hfill}{\settowidth\labelwidth{[999]}
        \leftmargin\labelwidth
        \advance\leftmargin\labelsep
        \@nmbrlisttrue\def\@listctr{pubctr}
        \setcounter{pubctr}{#1}\addtocounter{pubctr}{-1}}}
\def\@@publist{\list
        {[\arabic{pubctr}]\hfill}{\settowidth\labelwidth{[999]}
        \leftmargin\labelwidth
        \advance\leftmargin\labelsep
        \@nmbrlisttrue\def\@listctr{pubctr}}}
 \relax
\makeatother
\newskip\humongous \humongous=0pt plus 1000pt minus 1000pt

\newif\ifdtup

\relax

\def\be{\begin{equation}}
\def\ee{\end{equation}}
\def\ba{\begin{eqnarray}}
\def\ea{\end{eqnarray}}

\def\del{\partial}

\def\a{\alpha}

\def\b{\beta}

\def\g{\gamma}

\def\d{\delta}
\def\D{\Delta}

\def\om{\omega}

\def\l{\lambda}
\def\L{\Lambda}
\def\s{\sigma}

\def\cN{{\cal N}}

\def\cL{{\cal L}}

\def\no{\noindent}

\def\qq{\qquad}

\def\IR{\relax{\rm I\kern-.18em R}}

\def \ov {\over}

\def\IR{\relax{\rm I\kern-.18em R}}
\def\IL{\relax{\rm I\kern-.18em L}}

\def\inv{^{\raise.15ex\hbox{${\scriptscriptstyle -}$}\kern-.05em 1}}

\def\cL{{\cal L}}

\def\Tr{{\rm Tr}}

\begin{document}

\renewcommand{\theequation}{\thesection.\arabic{equation}}
\csname @addtoreset\endcsname{equation}{section}

\newcommand{\beq}{\begin{equation}}
\newcommand{\eeq}[1]{\label{#1}\end{equation}}
\newcommand{\ber}{\begin{equation}}
\newcommand{\eer}[1]{\label{#1}\end{equation}}
\newcommand{\eqn}[1]{(\ref{#1})}
\begin{titlepage}
\begin{center}

\hfill CPHT-RR002.012017

\vskip .3 in

{\large\bf Quantum aspects of doubly deformed CFTs}

\vskip 0.35in

{\bf G. Georgiou,$^1$\ \ E. Sagkrioti,$^2$\  \ K. Sfetsos}$^{2,3}$\ \ and\ \ {\bf K. Siampos}$^{4}$
\vskip 0.1in

\vskip 0.1in
{\em
${}^1$Institute of Nuclear and Particle Physics,\\ National Center for Scientific Research Demokritos,\\
Ag. Paraskevi, GR-15310 Athens, Greece
}
\vskip 0.1in

 {\em
${}^2$Department of Nuclear and Particle Physics,\\
Faculty of Physics, National and Kapodistrian University of Athens,\\
Athens 15784, Greece\\
}
\vskip 0.1in

{\em
${}^3$Centre de Physique Th\'eorique,\\
        Ecole Polytechnique, CNRS UMR 7644,\\
        Universit\'e Paris-Saclay,\\
        91128 Palaiseau Cedex, France
}

\vskip 0.1in

{\em${}^4$Albert Einstein Center for Fundamental Physics,\\
Institute for Theoretical Physics / Laboratory for High-Energy Physics,\\
University of Bern,
Sidlerstrasse 5, CH3012 Bern, Switzerland
}

\vskip 0.1in

{\footnotesize \texttt georgiou@inp.demokritos.gr, esagkrioti@phys.uoa.gr, ksfetsos@phys.uoa.gr, siampos@itp.unibe.ch}


\vskip .43in
\end{center}

\centerline{\bf Abstract}

\no
We study quantum aspects of the recently constructed doubly $\l$-deformed $\s$-models representing the effective
action of two WZW models interacting via current bilinears. We show that although the exact beta-functions and current anomalous dimensions are identical to those of the $\l$-deformed models,  this is not true for the anomalous dimensions of generic primary field operators in accordance with the fact that the two models differ drastically. Our proofs involve CFT arguments, as well as effective $\s$-model action and gravity calculations.

\vskip .4in
\noindent
\end{titlepage}
\vfill
\eject

\newpage

\tableofcontents

\noindent

\def\baselinestretch{1.2}
\baselineskip 20 pt
\noindent


\setcounter{equation}{0}
\section{Introduction  and conclusions}
\label{intro}
\renewcommand{\theequation}{\thesection.\arabic{equation}}
Recently, a new class of integrable $\s$-model theories based on current algebra theories for a general semisimple group G was constructed in \cite{Georgiou:2016urf}. This was achieved by utilizing a left-right asymmetric gauging of two independent WZW models both at the same positive integer level $k$, combined with two independent principal chiral models (PCM).
The resulting theories are characterized by $k$ and by two generic matrices $\l_1$ and  $\l_2$.
This class of theories was shown to be integrable for certain mono- or multi-parameter choices for $\l_1$ and  $\l_2$  \cite{Georgiou:2016urf}. A slight modification of the construction provides integrability for deformations corresponding to coset $G/H$ exact CFTs for cases in which $G/H$ is a symmetric space. The above construction, although different, is reminiscent of that for single $\l$-deformations \cite{Sfetsos:2013wia}.

The aim of this work is to study the quantum properties of the aforementioned models.
More precisely, we will derive the all-loop, exact in $\l_1$ and $\l_2$, but leading in the large $k$-expansion, $\b$-functions of the theory,
as well as the exact anomalous dimensions of the currents and of the bilinear operators that deform the CFT.
This will be achieved by using different independent methods which have their own advantages separately and in addition
they complement each other conceptually.
In the first one we employ CFT techniques to show that the correlation functions involving exclusively currents can be mapped to correlation functions of the single $\l$-deformed models calculated in \cite{Georgiou:2015nka, Georgiou:2016iom}.
We demonstrate that this can be done only for current correlators and not for generic correlation functions that involve affine primary operators.
In the second method, we employ the all-loop effective action of these doubly-deformed CFTs obtained
in \cite{Georgiou:2016urf} (provided in \eqref{defactigen} below) by considering  the Lagrangian of the quantum fluctuations around a classical constant background solution along the lines of \cite{Appadu:2015nfa} and for the case of isotropic coupling matrices, i.e. proportional to the identity. This method will provide the $\b$-functions for the two cases of principal interest, the isotropic double deformations corresponding to a general semisimple group $G$ and the double deformations corresponding to the symmetric coset space $G/H$ exact CFTs. In fact, these are actually the two cases for which this method is applicable. The results obtained are completely consistent with
those of the first method.
The gravitational background corresponding to the doubly $\l$-deformed is quite complicated. However, according to the findings of the first two methods we may compute the known
$\beta$-functions for the usual $\l$-deformed models by setting one of the matrices to zero identically. Indeed, we set $\l_2=0$ and use the $\b$-function of the resulting gravity background which in fact is quite simple. Then we determine the running of the remaining coupling $\l_1$ under the renormalization group (RG),
in the general case where $\l_1$ is an arbitrary matrix. For isotropic deformations the result is in complete agreement with that obtained from the two previous methods.

\no
We summarize the main results of the present paper: At the level of current operators the correlation functions of our model factorize and can be obtained from two copies each of which corresponds to a usual $\l$-deformed model, one with coupling matrix $\l_1$ and the other with coupling $\l_2$. In particular, this implies the remarkable fact that the running of each of the couplings $\l_1$ and $\l_2$ is the same as in the case of the single $\l$-deformations  computed in \cite{Sfetsos:2014jfa} and \cite{Itsios:2014lca} (see \eqref{b-func} and \eqn{dgdgg3}, respectively).
This is despite the fact that the doubly $\l$-deformed model is fundamentally different from the sum of two copies of
single $\l$-deformations. Furthermore, the anomalous dimensions of the currents and composite operators of the doubly deformed model \eqref{defactigen}
are given by \eqref{dim.current} and  \eqref{anom-comp}, respectively, in the large $k$-limit as computed in \cite{Georgiou:2015nka}.
Finally we calculate correlation functions involving affine primary operators by using CFT methods similar to the ones used in \cite{Georgiou:2016zyo,LeClair:2001yp}.
In this case the correlation functions will depend non-trivially on both $\l_1$ and $\l_2$ since they
have non-vanishing transformations with the left, as well as with the right currents.

Before proceeding, let us briefly review the models under consideration. The action defining
them depends on two group elements $g_i\in G,\,\,\,i=1,2$ and is given
by the deformation of the sum of two WZW models $S_{k}(g_1)$ and $S_{k}(g_2)$ \cite{Georgiou:2016urf}
\be
\begin{split}
&  S_{k,\l_1,\l_2}(g_1,g_2) = S_{k}(g_1) + S_{k}(g_2)
\\
&\qq\quad + {k\ov \pi} \int  \text{d}^2\s  \left(\!\! \begin{array}{cc}
    J_{1+}\! &\! J_{2+} \end{array}\!\!  \right)
\left(  \begin{array}{cc}
     \L_{21}\l_1 D_2^T\l_2 &   \L_{21}\l_1 \\
     \L_{12}\l_2 & \L_{12} \l_2 D_1^T\l_1\\
  \end{array} \right)
  \left(\!\! \begin{array}{c}
    J_{1-} \\ J_{2-} \end{array}\!\!\! \right) \ ,
\end{split}
\label{defactigen}
\ee
where the WZW action $S_{k}(g)$ for a group element $g\in G$  is given by
\be
\label{wzwacc}
S_{k}(g) = {k\ov 2\pi} \int \text{d}^2\s\, \Tr(\del_+ g^{-1} \del_- g)
+ {k\ov 12\pi} \int  \Tr(g^{-1} \text{d}g)^3\ .
\ee
and
\be
\L_{12}= (\mathbb{I} - \l_2 D_1^T \l_1 D_2^T)^{-1}\ ,\qq
\L_{21}= (\mathbb{I} - \l_1 D_2^T \l_2 D_1^T)^{-1}\ .
\ee
The matrices $D_{ab}$ and the currents $J^a_{\pm}$ are given by
\be
\label{hg3}
J^a_+ = - i\, \Tr(t^a \del_+ g g^{-1}) ,\qq J^a_- = - i\, \Tr(t^a g^{-1}\del_- g )\ ,
\qq D_{ab}= \Tr(t_a g t_b g^{-1})\  ,
\ee
where $t^a$'s are Hermitean matrices with $[t_a,t_b]=if_{abc}t_c$, here the structure constants $f_{abc}$ are taken to be
real.\footnote{\label{fsconv}The structure constants $f_{abc}$ are taken be real for gravity computations and imaginary for CFT ones, appearing in Secs. \ref{intro}, \ref{betaeffective}, \ref{betaRicci} and \ref{CFTcomput}, respectively.}  When a current or the matrix $D$ has an index $1$ or $2$ this means that one should use the corresponding group element in its definition.

\no
The above action has the following remarkable duality-type symmetry \cite{Georgiou:2016urf}\footnote{
The action is also invariant under the generalized parity transformation:
\begin{equation*}
\s^\pm\mapsto\s^\mp\,,\quad g_1\mapsto g_1^{-1}\,,\quad g_2\mapsto g_2^{-1}\,,\quad
\l_1\mapsto\l_2^T\,,\quad \l_2\mapsto\l_1^T\,,
\end{equation*}
similar to the $\l$-deformed action of \cite{Sfetsos:2013wia} found in \cite{Sfetsos:2014lla}.
}
\be
 k \mapsto -k \ , \quad \l_1 \mapsto  \l_1^{-1} \ ,\quad  \l_2 \mapsto  \l_2^{-1}\ ,
 \quad g_1\mapsto g_2^{-1}\ ,\quad g_2\mapsto g_1^{-1}\  .
\label{symmdual}
\ee
A similar non-perturbative duality is also present in the case of the $\l$-deformed action of \cite{Sfetsos:2013wia} as discovered in \cite{Itsios:2014lca,Sfetsos:2014jfa} and predicted before using path integral arguments in \cite{Kutasov:1989aw}.

\no
For small values of the entries of the matrices $\l_i$'s the action \eqn{defactigen} becomes
\be
\label{expan1}
S_{k,\l_1,\l_2}(g_1,g_2) = S_{k}(g_1) + S_{k}(g_2) + {k\ov \pi}
\int \text{d}^2\s\ \Big((\l_1)_{ ab} J^a_{1+} J^b_{2-} + (\l_2)_{ab} J^a_{2+} J^b_{1-}\Big) + \cdots \ ,
\ee
thus representing a current-current deformation of the original WZW actions. Notice, however, that
unlike the $\l$-deformed action \cite{Sfetsos:2013wia} the currents building the bilinear interactions belong to different
WZW models. The action \eqn{defactigen} is said to be the effective all-loop in $\l_1$ and $\l_2$ action corresponding to
\eqn{expan1} defined as a model on its own.


\section{Exact $\b$-functions, current \& primary fields correlators}
\label{CFTcomput}

In this section, we will calculate the exact expressions for the running of the couplings $\l_1$ and $\l_2$,
as well as for the anomalous dimensions of the currents $J^a_{1\pm}$,  $J^a_{2\pm}$ and current bilinears $J^a_{1+} J^b_{2-}$  and $J^a_{2+} J^b_{1-}$.
We will also derive the exact scaling dimensions of the affine primary operators of the model under consideration.

\subsection{Exact $\b$-functions \& current correlators}

Consider all the correlation functions involving current operators or composite current operators and split them into two sets. We will work in the Euclidean regime and
denote the Euclidean versions of $J_{i+}^a$ and $J_{i-}^a$, by $J_{i}^a$ and $\bar J_{i}^a$, respectively.
In the first set we assemble $J^a_{1}$,  $\bar J^a_{2}$ and all the composite operators built from these two and we do the same
for $J^a_{2}$, $\bar J^a_{1}$ and all of their composite operator, that is
\be
\label{operators}
{\mathcal O}=\{J^a_{1},\,\,\bar J^a_{2}, \,\,J^a_{1} \bar J^b_{2},\,\,\cdots\}\ ,
\qq { \tilde {\mathcal  O}}=\{J^a_{2},\,\,\bar J^a_{1}, \,\,J^a_{2} \bar J^b_{1},\,\,\cdots\}\ .
\ee
Our aim is to evaluate correlation functions involving an arbitrary number of ${\mathcal O}$ and ${ \tilde {\mathcal  O}}$, namely
\be
\langle \prod_{i=1}^n{\mathcal O}_{i}(z_i) \prod_{j=1}^m { \tilde{\mathcal O}}_{j}(z_j)\rangle= \frac{1}{\mathcal Z}\langle \prod_{i=1}^n{\mathcal O}_{i}(z_i) \prod_{j=1}^m { \tilde{\mathcal O}}_{j}(z_j) \text{e}^{-{1\ov \pi}
\int \text{d}^2z\ \big((\l_1)_{ab} J^a_{1} \bar J^b_{2} + (\l_2)_{ab} J^a_{2} \bar J^b_{1}\big)} \rangle_0\ ,
\label{corr-gen}
\ee
where the interaction is the leading term in the small $\l$ expansion given in \eqref{expan1}. The symbol $\langle \dots \rangle_0$ in the right hand side of
\eqref{corr-gen} denotes the average performed over the currents with the CFT action being  $S_{k}(g_1) + S_{k}(g_2)$.
The crucial observation is that the particular form of the interaction vertices in \eqref{expan1} leads to a factorization of the correlation function \eqref{corr-gen} since the operators $\mathcal O$ can be contracted only with the currents coming from the expansion of $\displaystyle \text{e}^{-{1\ov \pi}\int \text{d}^2z \ (\l_1)_{ab} J^a_{1} \bar J^b_{2} }$ and
similarly the operators ${ \tilde{\mathcal O}}$ can be contracted only with the currents coming from the expansion of the exponential  $\displaystyle \text{e}^{-{1\ov \pi}
\int \text{d}^2z \  (\l_2)_{ab} J^a_{2} \bar J^b_{1}}$.
In conclusion \eqref{corr-gen} can be written as follows
\be
\langle \prod_{i=1}^n{\mathcal O}_{i}(z_i) \prod_{j=1}^m { \tilde{\mathcal O}}_{j}(z_j)\rangle=\langle \prod_{i=1}^n{\mathcal O}_{i}(z_i) \rangle \cdot \langle \prod_{j=1}^m { \tilde{\mathcal O}}_{j}(z_j)\rangle \ ,
\label{corr-gen1}
\ee
where
\be
\begin{split}
&\langle \prod_{i=1}^n{\mathcal O}_{i}(z_i) \rangle= \frac{1}{\mathcal Z_1}\langle \prod_{i=1}^n{\mathcal O}_{i}(z_i)  \text{e}^{-{1\ov \pi}
\int \text{d}^2z\ (\l_1)_{ab} J^a_{1} \bar J^b_{2} } \rangle_0\ ,
\\
&\langle \prod_{j=1}^m{ \tilde{\mathcal O}}_{j}(z_j) \rangle= \frac{1}{\mathcal Z_2}\langle \prod_{j=1}^m{ \tilde{\mathcal O}}_{j}(z_j)  \text{e}^{-{1\ov \pi}
\int \text{d}^2 z\ (\l_2)_{ab} J^a_{2} \bar J^b_{1} } \rangle_0\ .
\end{split}
\label{corr-gen1a}
\ee
Thus we see that at the level of current operators the correlation functions of our model can be obtained from two copies each of which corresponds to a usual $\l$-deformed model, one with coupling $(\l_1)_{ab}$ and the other with coupling $(\l_2)_{ab}$. The above factorization of correlators is only true when restricted to correlation functions involving exclusively currents. When one or more affine primary operators are involved in the correlation function then such
a factorization will no longer be true (except for primaries transforming trivially under the left or the right current
algebras). This will be shown in the next subsection.
This is in accordance with the fact that the $\s$-model action \eqn{defactigen} entangles 
the group elements $g_1$ and $g_2$ non-trivially in such a way that its action can not be written 
as a sum of two $\l$-deformed models, with coupling matrices
$\l_1$ and $\l_2$, respectively.

The aforementioned factorization of \eqref{corr-gen1} implies that the $\beta$-functions for the couplings $\l_1$ and $\l_2$ are the same as in the single $\l$-deformed theory since the correlation functions from which they are derived involve only currents and as such they take the form of two copies of $\l$-deformed models. The above assertion is valid for all values of the deformation matrices $\l_1$ and $\l_2$ and to all order in the level $k$.
In particular, for the case of isotropic couplings these read
 \cite{Kutasov:1989dt,Gerganov:2000mt,Itsios:2014lca}
\be
\boxed{
\label{b-func}
\b_i:=\frac{\text{d}\l_i}{\text{d}t}=-\frac{c_G\l_i^2}{2k(1+\l_i)^2}\,,\quad t:=\ln\mu^2\,,\quad i=1,2
}\ ,
\ee
to leading order in the $1/k$-expansion and exactly in the deformation parameters. $c_G$ is the second Casimir of the adjoint representation defined by the structure constants of the group through
$f_{acd}f_{bcd}=-c_G\,\delta_{ab}$, here the structure constants are
imaginary (see footnote \ref{fsconv}).
Similarly, the pairs of currents  ($J^a_{1}, \bar J^a_{2}$) and
($J^a_{2}, \bar J^a_{1}$) acquire anomalous dimensions that depend only on $\l_1$ and $\l_2$, respectively. For the isotropic case their exact in $\l_1$ and $\l_2$  large $k$ expressions are given by \cite{Georgiou:2015nka, Georgiou:2016iom}
\be
\label{dim.current}
\boxed{\
\gamma^{(J_{1+})}=\gamma^{(J_{2-})}=\frac{c_G\lambda_1^2}{k(1-\lambda_1)(1+\lambda_1)^3}\ , \quad \gamma^{(J_{2+})}=\gamma^{(J_{1-})}=\frac{c_G\lambda_2^2}{k(1-\lambda_2)(1+\lambda_2)^3}
}\ ,
\ee
which are both positive.
Furthermore, for the composite operators deforming the sum of the two CFTs we have from \cite{Georgiou:2015nka} that
\be
\boxed{\
\gamma^{(J^a_{1+} J^a_{2-})} =-
{2 c_G\ov k} \frac{ \lambda_1 (1-\lambda_1(1-\lambda_1))}{(1-\lambda_1)(1+\lambda_1)^3}\ ,\quad
\gamma^{(J^a_{2+} J^a_{1-})} =-
{2 c_G\ov k} \frac{ \lambda_2 (1-\lambda_2(1-\lambda_2))}{(1-\lambda_2)(1+\lambda_2)^3}
}\ ,
\label{anom-comp}
\ee
which are both negative.

\no
The above considerations can be easily extended to the  left-right asymmetric cases where the levels of the four currents algebras involved are different. This can be done using the corresponding results for the left-right asymmetric $\l$-deformations in \cite{Georgiou:2016zyo}.

\subsection{Exact dimensions of primary operators}
In this subsection, we calculate the anomalous dimensions of affine primary operators along the lines of \cite{Georgiou:2016iom,Georgiou:2016zyo}.
We will verify the expectation that in this case the dimensions, as well as all correlators will depend on both couplings $\l_1$ and $\l_2$ and will not just be
what was found for the single $\l$-deformations in \cite{Georgiou:2016iom,Georgiou:2016zyo}.
The reason is that generic primary fields are sensitive under transformations from both the left and the right current algebras.
Such results might help in finding out the fate of these theories under the RG flow towards the IR, especially in the case of unequal levels.

\no
The CFT we are studying contains two kinds of affine primary fields belonging to the original WZW models and transforming under the corresponding left and right current algebras. From this point we focus on one of them, i.e. $\Phi_{i,i'}(z,\bar z)$ which
under the action of the currents $J_{1}^a$ and $\bar J_{1}^a$
transforms in the irreducible representations $R$ and $R'$, with the corresponding matrices being $t_a$ and $\tilde t_a$. Hence, we have for the indices labelling them
that $i=1,2,\dots , \dim R$ and $i'=1,2,\dots , \dim R'$. In addition these primaries are inert under
the action of the currents belonging to the second WZW model $J_2^a$ and $\bar J_2^a$.
The relevant OPE equations are
\begin{equation}
\label{jjj}
\begin{split}
& J_{1}^a(z) \Phi_{i,i'}(w,\bar w) = -{1\ov \sqrt{k}} {(t_a)_i{}^j \Phi_{j,i'}(w,\bar w)\ov z-w}\ ,
\\
&\bar J_{1}^a(\bar z) \Phi_{i,i'}(w,\bar w) = {1\ov \sqrt{k}} {(\tilde t_a)^{j'}{}_{i'} \Phi_{i,j'}(w,\bar w)\ov \bar z- \bar w}\  ,
\\
&J_{2}^a(z) \Phi_{i,i'}(w,\bar w) ={\rm regular}\ , \qquad \bar J_{2}^a(\bar z) \Phi_{i,i'}(w,\bar w) ={\rm regular}\ .
\end{split}
\end{equation}
Our conventions for the transformation matrices are
 $[t_a,t_b]=f_{abc}t_c$ and $[\tilde t_a,\tilde t_b]=f_{abc}\tilde t_c$, i.e. here $f_{abc}$ are taken to be
 imaginary.
At the conformal point, these affine primary fields are also Virasoro primaries with holomorphic and antiholomorphic dimensions given by \cite{Knizhnik:1984nr}
\begin{equation}
\D_R = {c_R\ov 2k+ c_G}\ ,\qq \bar \D_{R'} = {c_{R'}\ov 2k+ c_G}\ ,
\label{ddcft}
\end{equation}
where $c_R$, $c_{R'}$ and $c_G$ are the quadratic Casimir operators in the representations $R$, $R'$ and in the
adjoint representation. For the latter $(t_a)_{bc}= - f_{abc}$. They are defined as
\begin{equation}
(t_a t_a)_i{}^j = c_R \delta_i{}^j \ ,\qq (\tilde t_a \tilde t_a)_{i'}{}^{j'}
= c_{R'} \delta_{i'}{}^{j'}\ ,\qq  f_{acd} f_{bcd} = - c_G \delta_{ab} \ .
\end{equation}
Finally, the current OPEs are given in our conventions by
\be
J_1^a(z) J_1^a(z) ={\d_{ab}\ov (z-w)^2}+ {f_{abc} \ov \sqrt{k}} {J^c_1(w)\ov (z-w)}\,,
\ee
and similarly for the others.

\no
Next we proceed with the calculation of the exact dimensions of the primary fields.
A typical term in the perturbative expansion of the two-point function of primary fields will schematically have the form
$\l_1^n \l_2^m \langle \Phi^{(1)}_{i,i'}(x_1) \, (J^a_{1} \bar J^a_{2})^n\,  (J^b_{2} \bar J^b_{1})^m\,
\Phi^{(2)}_{j,j'}(x_2) \rangle$,
where $\Phi^{(1)}$ denotes an affine primary operator while $ \Phi^{(2)}$ denotes
its complex conjugate. The field $\Phi^{(1)}$ transfrorms as in \eqn{jjj} whereas
 $\Phi^{(2)}$ transforms similarly but with $t_a$ and $\tilde t_a$ replaced by
 $-t^*_a$ and $-\tilde t^*_a$, respectively. We will first argue that to order $1/k$ in the perturbative expansion that we are interested in, there will be no terms with both $n \neq 0$ and $m \neq 0$, that is, mixed terms of the two couplings $\l_1$ and $\l_2$ never appear.

\no
Consider the case when $n$ is an odd number.  We firstly choose to apply the Ward identity for one of the $\bar J^a_{2}$ currents. This current can not
be contracted with one of the external fields but only with another $\bar J^a_{2}$.
Once such contraction gives another $\bar J^a_{2}$ current via the non-Abelian part delivering another factor of $1/\sqrt{k}$. The remaining $\bar J^a_{2}$ currents, even in number in total, should then be
contracted among themselves only through the Abelian term of their OPE since in the opposite case the resulting expression will be of order higher or equal to $1/k^{3/2}$ in the large $k$-expansion and such terms are subleading. The next current for which we choose to apply the Ward identity is one of the $J^a_{1}$. This can be contracted either with another current of the same species through the non-Abelian term of the OPE or with one of the external fields. In both cases, we have already saturated the $1/k$ factor of the two-point function and as a result the currents associated with the second interaction term $(J^b_{2}
\bar J^b_{1})^m$ should be contracted only among themselves making the corresponding diagram disconnected.

\no
Now we turn to the case when $n$ is an even number and as before we apply first the
Ward identity for one of the currents $\bar J^a_{2}$ currents through the Abelian or the non-Abelian part of their OPE.
When this contraction is Abelian, at least one of the  $J^a_{1}$ currents should be contracted  with one of the external fields, otherwise all currents associated with
the first interaction vertex $(J^a_{1} \bar J^a_{2})^n $ will have been contracted among themselves giving a disconnected diagram. Thus, the contraction of the $J^a_{1}$ current with one of the external fields will leave us with an odd number of  $J^a_{1}$ currents which means that another $J^a_{1}$ current should be contracted with one of the external fields, hence saturating the factor $1/k$ to which we perform
computations. Hence, the currents associated with the second interaction term $(J^b_{2} \bar J^b_{1})^m$ have to be contracted only among themselves resulting to a disconnected diagram.
In the second case, when  two of the $\bar J^a_{2}$ currents are contracted through the non-Abelian part of their OPE, we are left with an odd number of $\bar J^a_{2}$ currents which means
that another non-Abelian contraction among two of the remaining $\bar J^a_{2}$ currents is necessary, hence saturating the $1/k$ overall factor. Then as before,  the currents associated with the second interaction vertices $(J^b_{2} \bar J^b_{1})^m$ should only be contracted among themselves giving a disconnected diagram.

\no
In conclusion, we have shown that the perturbative expansion can never produce terms containing mixed factors of $\l_1$ and  $\l_2$.
This implies that the anomalous dimension of the affine primary fields will take the generic form
\be
\label{prim-dim}
\g_{R,R'}={f_1(\l_1)c_R+f_2(\l_1)c_{R'}\ov k(1-\l_1)(1+\l_1)^3}+{h_1(\l_2)c_R+h_2(\l_2)c_{R'}\ov k (1-\l_2)(1+\l_2)^3}\ ,
\ee
where the pole structure of the dimensions in the above equation is dictated by the fact that each of the $\b$-functions of the model, $\b_{\l_1}$ and $\b_{\l_2}$,
is given by the same expression as in the single $\l$-deformed $\s$-model. Then one can use the Callan--Symanzik equation in a similar manner to that in \cite{Georgiou:2015nka}
in order to pin down the form of the anomalous dimensions.
The unknown polynomials $f_1$, $f_2$, $h_1$ and $h_2$ appearing in \eqref{prim-dim} can be determined by exploiting the symmetry of the action \eqref{symmdual}, as well as
the results from low order perturbation theory presented in the appendix.
From \eqn{fgjh} we have that
\be
\label{prim-dim-pert}
 \g_{R,R'}= {c_R \ov k}\left(1+ \l_1^2(1+\l_1^2)\right)+
 {c_{R'} \ov k}\l_2^2(1+\l_2^2)+ \mathcal O(\l^5/k)\ .
\ee
The symmetry of the model  \eqref{symmdual} when combined with the obvious $\mathbb{Z}_2$ symmetry exchanging $(g_1,\l_1)$ with $(g_2,\l_2)$ gives the following constraint for the
anomalous dimensions
\be\label{const}
\g_{R',R}(k,\l_1,\l_2)=\g_{R,R'}(-k,\l_2^{-1},\l_1^{-1})\ .
\ee

The exchange of the representations $R$ and $R'$ is related to the fact that under the combined symmetry mentioned 
above, the inversion of the group elements, i.e.
$g_i \mapsto g_i^{-1},\,i=1,2$, results into the interchange of the representations
$R$ and $R'$, for details see \cite{Georgiou:2016iom}.
Then \eqref{const} implies that the functions $f_1, f_2$ and $h_1, h_2$ obey the following relations
\be\label{const1}
f_1(\l_1)=h_2(\l_1^{-1}) \l_1^4\ , \qquad  f_2(\l_1)=h_1(\l_1^{-1}) \l_1^4\ .
\ee
Hence, the unknown polynomials are of order $\l_1^4$, at most. We fix them by in addition requiring agreement with the perturbative result \eqref{prim-dim-pert}. In fact, one needs the perturbative result only up to order $\l_1^2$ and $\l_2^2$.
In this way, we get $f_2=h_1=0$ and and $f_1(\l)=(1+\l)^2$ and $h_2(\l)=\l^2(1+\l)^2$.
Therefore the exact anomalous dimension is
\be\label{prim-dim-fin}
\boxed{
\g_{R,R'}=\frac{c_R}{k}\,{1\ov 1-\l_1^2}+\frac{c_{R'}}{k}\,{\l_2^2\ov 1-\l_2^2}
}\,.
\ee
This reproduces correctly the perturbative result \eqref{prim-dim-pert} up to order $\l_1^4$ and $\l_2^4$ and serves as a non-trivial check of the all-loop
expression \eqref{prim-dim-fin}\footnote{The exponent $\bar \g_{R,R'}$ of the $\bar x_{12}^2$ term in the 2-point function of primary fields is given by  \eqref{prim-dim-fin} but with $c_R$ and $c_R'$ exchanged.}. We see that, unlike the case with the current's anomalous dimensions,  it depends on both deformation parameters $\l_1$ and $\l_2$. In addition,
comparing with the anomalous dimensions of primary fields for the $\l$-deformed models \cite{Georgiou:2016iom} a main difference is the absence of a term proportional to the eigenvalues of the matrix $t_a\otimes  t_a^*$. Such a matrix does not
appear here. The reason lies on the fact that in the $\l$-deformed models the deformation is 
driven by current bilinears of the same WZW model, whereas here by current bilinears of different WZW models.

\section{Isotropic couplings: RG flows from the effective action}
\label{betaeffective}

In this section, we will employ the all-loop effective action of  \eqref{defactigen} in order to determine
the $\beta$-functions for the double isotropic deformation of two WZW models for a group $G$ as well as for two coset CFTs
for which $G/H$ is a symmetric space.

\subsection{Group space}

We first consider the case of two isotropic couplings for a group $G$ so that $(\l_i)_{ab} =\l_{i}\, \d_{ab}$, $i=1,2$.
To compute the $\beta$-functions we need to specify a classical background solution and compute the quantum fluctuations around it.
Of course self-consistency requires that the result is background independent.
The discussion goes along the lines of \cite{Appadu:2015nfa}.
The equations of motion for our models can be cast in the form \cite{Georgiou:2016urf}
\be
\label{eom.group}
\partial_\pm I^{i}_\mp=\mp\frac12\,[I^{i}_+,I^{i}_-]\ ,\qquad i=1,2\ ,
\ee
with
\be
\label{eom.group.rescale}
I^{i}_\pm=-\frac{2}{1+\l_{i}}\,A^{i}_\pm\ ,\qq i=1,2\ .
\ee
Consider group elements of the form\footnote{In what follows,
we denote by $\s^{\a}$, $\a=\pm$ the world-sheet coordinates.}
\be
\label{clas.sol.group}
g_i=\text{e}^{\sigma^\a\Theta^i_\a}\ ,\qquad
i=1,2\ ,
\ee
where the $\Theta^{i}_\a$'s
are arbitrary constant commuting elements of the Lie algebra $\mathfrak{g}$ of $G$.
The gauge fields evaluated at the above classical configuration
follow by inserting the classical solutions in the expressions for the gauge fields obtained in
eqs. (2.9) and (2.10) of \cite{Georgiou:2016urf} (with $A_\pm
\mapsto A^1_\pm$ and  $B_\pm \mapsto A^2_\pm$ to conform with the notation in the present paper)
\be
\label{rescale.group}
\begin{split}
&
A^{1}_+=\frac{\l_{1}}{1-\l_1\l_2}\left(\Theta^{1}_+ +\l_{2}\,\Theta^{2}_+\right)\ ,\quad
A^{1}_-=-\frac{\l_{1}}{1-\l_1\l_2}\left(\Theta^{2}_-+\l_{2}\,\Theta^{1}_-\right)\ .
\\
&
A^{2}_+=\frac{\l_{2}}{1-\l_1\l_2}\left(\Theta^{2}_+ +\l_{1}\,\Theta^{1}_+\right)\ ,\quad
A^{2}_-=-\frac{\l_{2}}{1-\l_1\l_2}\left(\Theta^{1}_-+\l_{1}\,\Theta^{2}_-\right)\ .
\end{split}
\ee
Therefore the $I^{i}_\pm$, for $i=1,2$ become constant commuting matrices which we denote by $I^{i}_{0,\pm}$
so that the equations of motion are indeed satisfied.
The Lagrangian evaluated on the background fields reads
\be
\label{Lag.back.group}
{\cal L}^{(0)} = -
{ k\ov 2\pi (1-\l_1\l_2)} \left(
                              \begin{array}{cc}
                               \Theta^1_+ & \Theta^2_+ \\
                              \end{array}
                            \right)\left(
                                                      \begin{array}{cc}
                                                    1+   \l_1\l_2& 2 \l_1 \\
                                                      2  \l_2 & 1+ \l_1\l_2 \\
                                                      \end{array}
                                                    \right)\left(
                                                             \begin{array}{c}
                                                               \Theta^1_- \\
                                                              \Theta^2_- \\
                                                             \end{array}
                                                           \right)\ .
\ee
To compute the  one-loop $\beta$-function we expand \eqref{eom.group} around the classical solution \eqref{clas.sol.group}
and we derive the operator acting on the fluctuations.
We let $I^{i}_\pm = I^{i}_{0,\pm} + \d I_\pm^{i}$ and
we linearize the equations of motion \eqn{eom.group}.
After some rearrangements we get that
\be
{\mathcal D}^{i}\left(
                              \begin{array}{c}
                               \delta I^{i}_+ \\ \delta I^{i}_- \\
                              \end{array}
                            \right)=0\ ,\qq i=1,2\ ,
\ee
where the matrix differential operators acting on the fluctuations are given by
\be
{\mathcal D}^{i}=\left(
                                                      \begin{array}{cc}
                                                        \partial_--\frac{1}{2}\tilde I^{i}_-& \frac{1}{2}\tilde I^{i}_+ \\
                                                        \frac{1}{2}\tilde I^{i}_- &  \partial_+-\frac{1}{2}\tilde I^{i}_+ \\
                                                      \end{array}
                                                    \right)\ , \quad i=1,2\ ,
                                                    \ee
with
\be
\label{current.adjoint.group}
\left(\tilde I^{i}_\pm\right)_{ab}=if_{abc}\,I^{i}_{0,\pm c}\ ,\quad i=1,2\ .
\ee
The one-loop effective Lagrangian in momentum space, after Wick rotating to Euclidean space and integrating out the fluctuations
appearing in a Gaussian path integral, reads\footnote{The analytic continuation $\left(\tau,\sigma\right)\mapsto\left(i\sigma_1,\sigma_2\right)$ and so $(p_0,p_1)\mapsto(ip_1,p_2)$.}
\be
\label{Lag.eff.group}
-{\cal L}^\text{eff}_\text{E}={\cal L}^{(0)}+\int^\mu\frac{\text{d}^2p}{(2\pi)^2}\ln\det
\left(
                                                      \begin{array}{cc}
                                                        {\widehat{\mathcal D}}^{1}& {\bf 0} \\
                                                        {\bf 0} &  \widehat{{\mathcal D}}^{2} \\
                                                      \end{array}
                                                    \right)^{-1/2}\ ,
\ee
where
\be
\widehat{{\mathcal D}}^{i}=\left(
                                                      \begin{array}{cc}
                                                        p_--\frac{1}{2}\tilde I^{i}_-& \frac{1}{2}\tilde I^{i}_+ \\
                                                        \frac{1}{2}\tilde I^{i}_- &  p_+-\frac{1}{2}\tilde I^{i}_+ \\
                                                      \end{array}
                                                    \right)\,,\quad i=1,2\ ,\quad p_\pm=\frac{1}{2}\left(p_1\pm i\, p_2\right)\ .
 \ee
Next we evaluate the determinant in \eqref{Lag.eff.group}
\be
\ln\det
\left(
                                                      \begin{array}{cc}
                                                        {\widehat{\mathcal D}}^{1}& {\bf 0} \\
                                                        {\bf 0} &  \widehat{{\mathcal D}}^{2} \\
                                                      \end{array}
                                                   \right)=
                                                   \ln\det{\widehat{\mathcal D}}^{1}+  \ln\det{\widehat{\mathcal D}}^{2}\,,
 \ee
where
\be
\begin{split}
& \ln\det{\widehat{\mathcal D}}^{i}=
 \ln\det C+\text{Tr}\ln\left(\mathbb{I}_2+C^{-1}E_{i}\right)\,,\\
 &C=  \left(\begin{array}{cc}
                                                       p_-& 0 \\
                                                        0 &  p_+ \\
                                                      \end{array}\right)\,,\quad
                                                      E_{i}=\frac12\,\left(\begin{array}{cc}
                                                      -\tilde I^{i}_-& \tilde I^{i}_+ \\
                                                        \tilde I^{i}_- & -\tilde I^{i}_+ \\
                                                      \end{array}\right)\ ,\quad i=1,2\  .
 \end{split}
\ee
To proceed we expand the field dependent term
\be
\text{Tr}\ln\left(\mathbb{I}_2+C^{-1}E_{i}\right)=-\frac{\text{Tr}\left(\tilde I^{i}_+p_-+\tilde I^{i}_-p_+\right)^2}{2p^4}
+{\cal O}\left(\frac{p_\pm^4}{p^8}\right)\ ,\quad i=1,2\ .
\ee
The logarithmically divergent term in  \eqref{Lag.eff.group} comes only from the explicitly
depicted term above. After performing the momentum integration we get that
\be
\label{Lag.eff.group.final}
\begin{split}
& {\cal L}^\text{eff}_\text{E}=-{\cal L}^{(0)}-\frac{1}{4\pi}\text{Tr}\left(\tilde I^1_+\tilde I^1_-+\tilde I^2_+\tilde I^2_-\right)\,\ln\mu
\\
& \phantom{xxx} =
-{\cal L}^{(0)}-\frac{c_G}{4\pi}\left( I^{1}_{0,+a}  I^{1}_{0,-a}+  I^{2}_{0,+a}  I^{2}_{0,-a}\right)\,\ln\mu\ ,
\end{split}
\ee
where we have used \eqref{current.adjoint.group} to obtain the second line above.
The one-loop $\beta$-function is derived by
demanding that the effective action \eqref{Lag.eff.group.final}, after inserting \eqref{eom.group.rescale}, \eqref{rescale.group}
and \eqref{Lag.back.group} is independent of the cutoff scale $\mu$. After some algebraic manipulations we obtain the same
result as that from
the field theory calculation \eqref{b-func}.
This agreement is non-trivial evidence that \eqref{defactigen}
is indeed the all-loop effective action of the linearized model described by \eqref{expan1}.

\subsection{Symmetric space}

Consider now the case of a deformation of two coset CFTs corresponding to symmetric spaces.
For convenience we spit the group index into a part belonging to the subgroup $H$ and the rest
corresponding to the coset. We denote by Latin letters the subgroup indices and by Greek letters
those of the coset. Using this notation the coupling matrices have non-vanishing elements \cite{Georgiou:2016urf}
\be
(\l_i)_{ab} =\delta_{ab}\,,\quad (\l_i)_{\a\b}=\l_{i}\,\delta_{\a\b}\ ,\quad i=1,2\ .
\ee
We also split the fields in the subgroup and coset components as
\be
I^{i}_\pm=I_\pm^{h|i}+I_\pm^{g/h|i}\ ,\qq i=1,2\ .
\ee
The equations of motion are of the form \cite{Georgiou:2016urf}
\be
\label{eom.coset}
\begin{split}
&\partial_+I_-^{h|i}-\partial_-I_+^{h|i}+[I_+^{h|i},I_-^{h|i}]+[I_+^{g/h|i},I_-^{g/h|i}]=0\ ,
\\
&\partial_\pm I_\mp^{g/h|i}=[I_\mp^{g/h|i},I_\pm^{h|i}]\ ,\qq i =1,2\ ,
\end{split}
\ee
with
\be
\label{eom.coset.rescale}
I_\pm^{h|i}=-A_\pm^{h|i}\,,\quad I_\pm^{g/h|i}=-\frac{A_\pm^{g/h|i}}{\sqrt{\l_{i}}}\ , \quad i=1,2
\ee
and we have used the fact that $f_{\a\b\g}=0$ for symmetric spaces.
We fix the residual gauge by enforcing the covariant gauge fixing condition
\be
\partial_+I_-^{h|i}+\partial_-I_+^{h|i}=0\ , \qq i =1,2\ .
\label{gfcoset}
\ee
We will comment on other gauge choices at the end of  this section.
As in the group case, to derive the $\b$-function we need to identify the proper background classical solution.
We make the same choice as in \eqn{clas.sol.group} but now $\Theta^{i}_\a$, $i=1,2$
are arbitrary constant commuting elements of
$\mathfrak{g}/\mathfrak{h}$.
The Lagrangian $\cL^{(0)}$ evaluated on the background fields has the same form \eqn{Lag.back.group}. In addition the gauge fields in the
coset $A_\pm^{g/h|1}$ and $A_\pm^{g/h|2}$ take the form of \eqn{rescale.group} whereas those in the subgroup $A_\pm^{h|1}=A_\pm^{h|2}=0$.

\no
Varying the equations of motion \eqn{eom.coset} and the gauge fixing condition \eqn{gfcoset} yields
\be
{\mathcal D}^{i}\left(
                              \begin{array}{c}
                               \delta I^{g/h|i}_+ \\ \delta I^{g/h|i}_- \\
                               \delta I^{h|i}_+ \\ \delta I^{h|i}_- \\
                              \end{array}
                            \right)=0\ , \qq i =1,2\ ,
\ee
where
\be
\label{fdkk11}
{\mathcal D}^{i}=\left(
                                                      \begin{array}{cccc}
                                                      \partial_-& 0 & 0 & \widetilde I^{g/h|i}_+\ \\
                                                        0 & \partial_+ & \widetilde I^{g/h|i}_- &0 \\
                                                        \widetilde I^{g/h|i}_-& -\widetilde I^{g/h|i}_+&-\partial_-&\partial_+\\
                                                        0&0&\partial_-&\partial_+\\
                                                      \end{array}
                                                    \right)\ , \qq i =1,2\ ,
                                                    \ee
with
\be
\label{current.adjoint.coset}
\left(\tilde I^{g/h|i}_\pm\right)_{\a b}=if_{\a b\g}\,I^{g/h|i}_{\pm \g}\ , \qq i =1,2\ .
\ee
The one-loop effective Lagrangian in momentum space, after Wick rotating to Euclidean space takes the form
\eqn{Lag.eff.group} where now
\be
\label{dshfkh}
\widehat{{\mathcal D}}^{i}=\left(
                                                      \begin{array}{cccc}
                                                      p_-& 0 & 0 & \widetilde I^{g/h|i}_+\ \\
                                                        0 & p_+ & \widetilde I^{g/h|i}_- &0 \\
                                                        \widetilde I^{g/h|i}_-& -\widetilde I^{g/h|i}_+&-p_-&p_+\\
                                                        0&0&p_-&p_+\\
                                                      \end{array}
                                                    \right)\ , \qq i =1,2\ .
 \ee
Working along the lines with the group case we get
\be
\label{Lag.eff.coset.final}
\begin{split}
&{\cal L}^\text{eff}_\text{E}=-{\cal L}^{(0)}-\frac{1}{\pi}\text{Tr}\left( \tilde I^{g/h|1}_{+}  \tilde I^{g/h|1}_{-}+ \tilde  I^{g/h|2}_{+} \tilde I^{g/h|2}_{-}\right)\,\ln\mu\,,
\\
&
\phantom{xxx} =-{\cal L}^{(0)}-\frac{c_G}{\pi}\left(  I^{g/h|1}_{+\a}   I^{g/h|1}_{-\a}+   I^{g/h|2}_{+\a}  I^{g/h|2}_{-\a}\right)\,\ln\mu\ ,
\end{split}
\ee
due to \eqref{current.adjoint.coset} and the fact that for symmetric spaces $f_{\a\b\g}=0$.

\no
As before the one-loop $\beta$ function can be derived by
demanding that the effective action \eqref{Lag.eff.coset.final}
is independent of the cutoff scale $\mu$. The result is
\be
\label{cossst}
\boxed{\
\beta_i=-\frac{c_G\l_i}{2 k}\quad i=1,2
}\ ,
\ee
for arbitrary constant $\Theta^{i}_\a$'s. This result is identical to
that obtained for the $\l$-deformed $SU(2)/U(1)$ coset using gravity in \cite{Itsios:2014lca} and generalized for $\l$-deformations for arbitrary $G/H$ symmetric coset CFTs in \cite{Appadu:2015nfa}.
We end this section by noting that a different gauge choice than \eqn{gfcoset} (necessarily involving only $A_\pm^{h|{i}}$) would have resulted in different (43) and (44) elements in \eqn{fdkk11} and \eqn{dshfkh}. It turns out that this does not affect the logarithmic behaviour in \eqn{Lag.eff.coset.final}.
Hence \eqn{cossst} is unchanged as it should be.


\section{A simple action and the $\beta$-function from gravity}
\label{betaRicci}

In this section, we consider the special case where $\l_2=0$.
The other matrix $\l_1$, renamed as $\l$, will be kept general.
We will use the expressions for the running of the $\s$-model couplings given in terms of the metric and antisymmetric tensor fields of \cite{honer,Friedan:1980jf,Curtright:1984dz} in order to determine the running of the couplings $\l_{ab}$. The resulting expression is in complete agreement with the CFT we have provided and will coincide with the general result for the $\l$-deformed backgrounds
computed in \cite{Sfetsos:2014jfa}. However, in the case at hand the computation will be significantly easier since the corresponding
action extremely simplifies and reads
\be
\label{acl2miden}
S_{k,\l}(g_1,g_2) = S_{k}(g_1) + S_{k}(g_2)
+ {k\ov \pi} \int  \text{d}^2\s\, \l_{ab} J_{1+}^a J_{2-}^b \ ,
\ee
which we emphasize 
is not an approximation for small entries $\l_{ab}$, but it is just obtained from \eqn{defactigen} as described above.

\subsection{Computation of the $\beta$-function}

To proceed with the computation we first read off the line element
\be
\text{d}s^2 = R^a R^a + L^{\hat a} L^{\hat a} + 2\l_{ab} R^a L^{\hat b}\ ,
\ee
where
\be
\begin{split}
&R^a= - i\, \Tr(t^a \text{d} g_1 g_1^{-1})\ ,\qq L^{\hat a} = - i\, \Tr(t^a g_2^{-1}\text{d} g_2 )\ ,
\\
&\text{d}L^a=\frac12\,f_{abc}L^b\wedge L^c\,,\quad \text{d}R^a=-\frac12\,f_{abc}R^b\wedge R^c\ .
\end{split}
\ee
Hence, the unhatted and hatted indices denote the Maurer--Cartan forms of $g_1$ and $g_2$ respectively.
By introducing the vielbeins
\be
\label{gen.vielbein}
\text{e}^a =R^a\ ,\quad \text{e}^{\hat a} = L^{\hat a}
+ \l_{ba}R^b\,,
\ee
as well as the double index notation $A=(a,\hat a)$ the line element can be written as
\be
\text{d}s^2 = \tilde g_{ab}\text{e}^a\text{e}^b+\text{e}^{\hat a}\text{e}^{\hat a}=G_{AB}\,
\text{e}^A\text{e}^B\ ,
\ee
where $\tilde g_{ab}=(\mathbb{I}-\l\l^T)_{ab}$ and
for later use we also define $g_{ab}=(\mathbb{I}-\l^T\l)_{ab}$.
We will also need the two-form which is given by
\be
B =B_0 +  \l_{ab}  R^a\wedge L^{\hat b}\,,
\ee
where $B_0$ is the two-form corresponding to the two WZW models with
\be
H_0 =\text{d}B_0= -{1\ov 6} f_{abc}\left( R^{ a}\wedge R^{ b}\wedge R^{c}+
 L^{\hat a}\wedge L^{\hat b}\wedge L^{\hat c}\right)\,.
\ee
Note that we have not included an overall factors of $\frac{k}{2\pi}$ in the definitions of $\text{d}s^2$ and $B$.
Also, the sign of $H_0$ is in accordance with that of the WZ term in \eqn{wzwacc}.

\no
The tangent metric $G_{AB}$ is constant and so the spin connection $\omega_{AB}$ is antisymmetric.
A practical way to compute it is by first defining the quantities
$C^A{}_{BC}= - C^A{}_{CB}$ from
\be
\text{d}\text{e}^A=\frac{1}{2}\,C^A{}_{BC}\,\text{e}^B\wedge\text{e}^C\ ,\quad C_{ABC}=G_{AD}\,C^D{}_{BC}\ .
\ee
Then, simply
\be
\omega_{AB}= \omega_{AB|C}\text{e}^C= \frac12\,\left(C_{ABC}-C_{CAB}+C_{BCA}\right) \text{e}^C\ ,
\ee
which also defines the useful, in explicit computations, quantity $\omega_{AB|C}$.
Employing the above and \eqref{gen.vielbein} we find that
\be
\begin{split}
&\omega_{ab}=-\frac{1}{2}\left(\tilde{g}_{ad}f_{dbc}-\tilde{g}_{cd}f_{dab}+\tilde{g}_{bd}f_{dca}\right)\,\text{e}^c
+\frac12\left(\l_{dc}f_{dab}-\l_{ad}\l_{be}f_{cde}\right)\,\text{e}^{\hat c}\,,\\
&\omega_{\hat a b}=\frac{1}{2}\,\left(f_{ade}\l_{bd}\l_{ce}-\l_{da}f_{dbc}\right)\,\text{e}^c\,,
\\
&
\omega_{\hat a\hat b}=-f_{abd}\l_{cd}\,\text{e}^c+\frac12\,f_{abc}\,\text{e}^{\hat c}\,.
\end{split}
\ee
Next we evaluate the field strength of the two-form
\be
\begin{split}
H=\text{d}B=&\left(-\frac16\,f_{abc}-\frac13\,\l_{ad}\l_{be}\l_{cf}f_{def}+\frac12\,f_{abd}\left(\l\l^T\right)_{cd}\right)\text{e}^a\wedge\text{e}^b\wedge\text{e}^c\\
&+\frac12\,\left(f_{ade}\l_{bd}\l_{ce}-\l_{da}f_{dbc}\right)\text{e}^{\hat a}\wedge \text{e}^b\wedge \text{e}^c-
\frac16\,f_{abc}\,\text{e}^{\hat a}\wedge\text{e}^{\hat b}\wedge\text{e}^{\hat c}\,.
\end{split}
\ee
It will be convenient to use spin connections which include the torsion, defined as
\be
\omega^\pm_{AB}=\omega_{AB}\pm\frac12\,H_{ABC}\,\text{e}^C = \om^\pm_{AB|C} \text{e}^C \ .
\ee
The torsion-full  spin connections are found to be
\be
\label{gen.spin.connec}
\begin{split}
&\omega^+_{ab}=\left(-f_{abc}-\l_{ad}\l_{be}\l_{cf}f_{def}+\left(\l\l^T\right)_{ad}f_{dbc}
+\left(\l\l^T\right)_{bd}f_{adc}\right)\,\text{e}^c\ ,
\\
&\omega^+_{\hat a b}=\left(f_{ade}\l_{bd}\l_{ce}-\l_{da}f_{dbc}\right)\,\text{e}^c\ ,
\\
& \omega^+_{\hat a\hat b}=-f_{abd}\l_{cd}\,\text{e}^c\
\end{split}
\ee
and
\be
\begin{split}
&\omega^-_{ab}=\left(\l_{ad}\l_{be}\l_{cf}f_{def}-\left(\l\l^T\right)_{cd}f_{abd}\right)\,\text{e}^c+
\left(\l_{dc}f_{dab}-\l_{ad}\l_{be}f_{dec}\right)\,\text{e}^{\hat c}\,,\\
&\omega^-_{\hat a b}=0\ ,
\\
& \omega^-_{\hat a\hat b}=-f_{abd}\l_{cd}\,\text{e}^c+f_{abc}\,\text{e}^{\hat c}\,.
\end{split}
\ee
Finally, we compute the torsion-full Ricci tensor by a useful rewriting
\be
\label{identity.Ricci}
R^\pm_{AB}=\partial_C\omega^{\pm C}{}_{A|B}-\omega^\pm_{AC|D}\omega^\mp_B{}^{D|C}-\nabla^\pm_B\omega^{\pm C}{}_{A|C}\,,
\ee
where $\partial_A=\text{e}_A{}^M\partial_M$.
The last term corresponds to a diffeomorphism associated with
\be
\omega^{\pm C}{}_{A|C}=\left(\partial_M\text{e}^C_N-\partial_N\text{e}^C_M\right)\text{e}^M_A\text{e}^N_C\ ,
\ee
which shows that it is a vector in target space.
The one-loop RG flow equations read
\be
{\text{d}\ov \text{d}t}(G_{MN}+B_{MN}) =R^-_{MN} + \nabla^+_N \xi_M \ ,
\ee
where the second term corresponds to diffeomorphisms along $\xi^M$.
The above may be rewritten in the tangent frame $\text{e}^A=\text{e}^A_M\,\text{d}X^M$ as
\be
\label{one.loop.frame}
{\text{d}\ov \text{d}t}(G_{MN}+B_{MN})=
\left(R^-_{AB}+\nabla_B^-\xi_A\right)\text{e}^A_M\text{e}^B_N
\ .
\ee
To proceed we evaluate the left hand side as
\be
{\text{d}\ov \text{d}t}(G_{MN}+B_{MN})
={\text{d}\lambda_{ab}\ov\text{d}t}\,R_M^a L_N^{\hat b}
={\text{d}\lambda_{ab}\ov\text{d}t}\,
\text{e}_M^a(\text{e}_N^{\hat b}-\lambda_{cb}\text{e}_N^c)
 \ ,
\ee
where we have used \eqref{gen.vielbein}. In addition we use the freedom to perform
diffeomorphisms in order to absorb in the expression for $R^{-}_{AB}$ the term involving
$\om^{-C}{}_{A|C}$, by choosing $\xi_A=\omega^{- C}{}_{A|C}$. Then,
employing the latter, as well as \eqref{gen.spin.connec}-\eqref{identity.Ricci}
and \eqref{one.loop.frame} we find that
\be
\boxed{\
{\mathrm{d}\l_{ab} \ov \mathrm{d}t} = {1\ov 2k} \cN(\l)_{ac}{}^d \cN(\l^T)_{bd}{}^c
\label{dgdgg3} \
} \ \ ,
\ee
with
\be
{\cN}(\l)_{ab}{}^c = \left(\l_{ae}\l_{bd}f_{edf} - f_{abe} \l_{ef}\right) g^{fc} \ ,
\ee
where we have reinstalled the overall integer $k$, which, it does not flow.
The expression \eqn{dgdgg3} is identical to that found in \cite{Sfetsos:2014jfa}. For the isotropic coupling case it reduces
to \eqn{b-func}, identical to the one of \cite{Itsios:2014lca}.

\subsection{Some properties of the action}

According to our general discussion when we consider the
action \eqn{acl2miden} the currents $J_{2+}^a $ and $J_{1-}^a $ acquire no anomalous dimension.
That implies that the action \eqn{acl2miden} should have on-shell chiral and anti-chiral currents.
The equations of motion corresponding to the variation of
the two group elements $g_1$ and $g_2$ can be readily derived.
They are most easily obtained from eqs. (3.6) and (3.7) of  \cite{Georgiou:2016urf} after using eqs. (2.9) and (2.10) of
the same work, setting $\l_2=0$ and renaming $\l_1$ by $\l$. Following this approach we obtain
\be
\label{eqqq1}
\begin{split}
&\l \del_+ J_{2-} + \del_- J_{1+} = i [J_{1+}, \l J_{2-}]\ ,\\
&\del_+ J_{2-} + \l^T \del_- J_{1+} = i [\l^T J_{1+}, J_{2-}]\
\end{split}
\ee
and
\be
\label{eqqq2}
\begin{split}
& \del_- {\cal J}_+ = 0 \ ,\qq {\cal J}_+ =   J_{2+}  + D_2 \l^T J_{1+}\ ,
\\
& \del_+ {\cal J}_- = 0 \ ,\qq {\cal J}_- = J_{1-}  + D_1^T \l J_{2-}\ .
\end{split}
\ee
The first (second) of \eqn{eqqq1}
is equivalent to the second (first) of \eqn{eqqq2}. To prove that
we used the identities $(D^T \del_- D)^{ab} = f^{ab}{}_c J^c_-$ and $(\del_+ D D^T)^{ab} = f^{ab}{}_c J^c_+$.
The above chiral and anti-chiral conserved currents ${\cal J}_\pm$ are
deformations of $J_{2+}$ and $J_{1-}$,
to which they reduce for $\l=0$. This is consistent with their
vanishing anomalous dimensions.  The equations \eqn{eqqq2} for the action \eqn{acl2miden} were derived before in
\cite{Hull:1995gj}.

\subsection*{Acknowledgements}

K. Siampos' work was partially supported by the \textsl{Germaine de Sta\"el} Franco--Swiss bilateral
 program 2015 (project no 32753SG). G. Georgiou and K. Siampos would like to thank
 the Physics Department of the National and Kapodistrian University of Athens for hospitality.


\appendix

\section{Perturbative 2-point functions for primary fields}

In this appendix we will compute the $\langle\Phi^{(1)}_{i,i'}\Phi^{(2)}_{j,j'}\rangle$ correlator up to
four-loop $\mathcal{O}(\l^4)$ and to leading order in the large $k$ expansion. We concentrate
to the case where these fields transform non-trivially under the $J_1^a$ and $\bar J_1^a$ as in \eqn{jjj}.

\no
\textbf{One-loop} $\mathcal{O}(\l)$: It is easily seen that the corresponding contribution
is zero, i.e.
\begin{equation}
\label{FF1}
\langle\Phi^{(1)}_{i,i'}(x_1,\bar{x}_1)\Phi^{(2)}_{j,j'}(x_2,\bar{x}_2)\rangle^{(1)}_{\l}=0\ .
\end{equation}

\no
\textbf{Two-loop} $\mathcal{O}(\l^2)$: To this order we find that
\begin{equation}
\label{a22}
\langle\Phi^{(1)}_{i,i'}(x_1,\bar{x}_1)\Phi^{(2)}_{j,j'}(x_2,\bar{x}_2)\rangle^{(2)}_{\l}=
\frac{1}{2!\pi^2}\int \text{d}^2z_{12} \left(\l_1^2\,A_2+\l_2^2\,B_2\right) ,
\end{equation}
where  
\begin{equation}
\begin{split}
& A_2=\langle\Phi^{(1)}_{i,i'}(x_1,\bar{x}_1)J_1^{a_1}(z_1)\bar{J}_2^{a_1}(\bar{z}_1)J_1^{a_2}(z_2)\bar{J}_2^{a_2}(\bar{z}_2)\Phi^{(2)}_{j,j'}(x_2\ ,\bar{x}_2)\rangle\,,
\\
&
 B_2=\langle\Phi^{(1)}_{i,i'}(x_1,\bar{x}_1)J_2^{a_1}(z_1)\bar{J}_1^{a_1}(\bar{z}_1)J_2^{a_2}(z_2)\bar{J}_1^{a_2}(\bar{z}_2)\Phi^{(2)}_{j,j'}(x_2\ ,\bar{x}_2)\rangle\ .
 \end{split}
\end{equation}
We note that the mixed terms proportional to  $\l_1\l_2$ do not contribute, as explained
in the main text.
After further contractions with $J_1^{a_1}$, followed by contractions of $\bar{J}_2^{a_1}$ with $\bar{J}_2^{a_2}$ we end up with
\be
\begin{split}
& A_2=\frac{c_R}{k}\frac{(\mathbb{I}_R\otimes\mathbb{I}_{R'})_{ii',jj'}}{x_{12}^{2\Delta_R}\bar{x}_{12}^{2\bar{\Delta}_{R'}}}\left(\frac{1}{\bar{z}_{12}^2(z_1-x_1)(z_2-x_1)}-\frac{1}{\bar{z}_{12}^2(z_1-x_1)(z_2-x_2)}\right.
\\
&\qq\qq\qq\qq\qq \phantom{x}\left.
-\frac{1}{\bar{z}_{12}^2(z_1-x_2)(z_2-x_1)}+\frac{1}{\bar{z}_{12}^2(z_1-x_2)(z_2-x_2)}   \right)\ .
\end{split}
\ee
The expression for $B_2$ is found by performing a parity transformation in $A_2$ to be
\be
\begin{split}
& B_2=\frac{c_{R'}}{k}\frac{(\mathbb{I}_R\otimes\mathbb{I}_{R'})_{ii',jj'}}{x_{12}^{2\Delta_R}\bar{x}_{12}^{2\bar{\Delta}_{R'}}}\left(\frac{1}{z_{12}^2(\bar{z}_1-\bar{x}_1)(\bar{z}_2-\bar{x}_1)}-\frac{1}{z_{12}^2(\bar{z}_1-\bar{x}_1)(\bar{z}_2-\bar{x}_2)}\right.
\\
&\left.\qq\qq\qq\qq\qq\phantom{x} -\frac{1}{z_{12}^2(\bar{z}_1-\bar{x}_2)(\bar{z}_2-\bar{x}_1)}+\frac{1}{z_{12}^2(\bar{z}_1-\bar{x}_2)(\bar{z}_2-\bar{x}_2)}   \right)\ .
\end{split}
\ee
Then we perform the double integration in \eqn{a22}, choosing the order of integration from left to right.  Using the symmetry under $x_1\leftrightarrow x_2$ we compute the final result for the two-point function of primary fields up to order $\mathcal{O}(\l^2/k)$
\begin{equation}
\label{FF2}
\langle\Phi^{(1)}_{i,i'}(x_1,\bar{x}_1)\Phi^{(2)}_{j,j'}(x_2,\bar{x}_2)\rangle^{(2)}_{\l}={1\ov k}\big(c_R\l_1^2+c_{R'}\l_2^2\big)\frac{(\mathbb{I}_R\otimes \mathbb{I}_{R'})_{ii',jj'}}{x_{12}^{2\D_R}\bar{x}_{12}^{2\bar{\D}_{R'}}}
\ln\frac{\varepsilon^2}{|x_{12}|^2}
\ .
\end{equation}
All necessary integrals have been encountered before in similar computations in various
works, for example in \cite{Georgiou:2016iom}.

\no
\textbf{Three-loop} $\mathcal{O}(\l^3)$: To this order, we have that
\begin{equation}
\langle\Phi^{(1)}_{i,i'}(x_1,\bar{x}_1)\Phi^{(2)}_{j,j'}(x_2,\bar{x}_2)\rangle^{(3)}_{\l}=
-\frac{1}{3!\pi^3}\int \text{d}^2z_{123} (\l_1^3A_3 + \l_2^3B_3)\,,
\end{equation}
where
\begin{equation}
\begin{split}
&A_3= \langle\Phi^{(1)}_{i,i'}(x_1,\bar{x}_1)J_1^{a_1}(z_1)\bar{J}_2^{a_1}(\bar{z}_1) J_1^{a_2}(z_2)\bar{J}_2^{a_2}(\bar{z}_2) J_1^{a_3}(z_3)\bar{J}_2^{a_3}(\bar{z}_3) \Phi^{(2)}_{j,j'}(x_2,\bar{x}_2)\rangle\ ,
\\
& B_3= \langle\Phi^{(1)}_{i,i'}(x_1,\bar{x}_1)J_2^{a_1}(z_1)\bar{J}_1^{a_1}(\bar{z}_1) J_2^{a_2}(z_2)\bar{J}_1^{a_2}(\bar{z}_2) J_2^{a_3}(z_3)\bar{J}_1^{a_3}(\bar{z}_3) \Phi^{(2)}_{j,j'}(x_2,\bar{x}_2)\rangle\ .
\end{split}
\end{equation}
As explained in the main text, terms with mixed factors of $\l_1$ and $\l_2$ do not occur.
Furthermore, under a parity transformation, mapping
$J_i\leftrightarrow\bar{J}_i$ and $z_i\leftrightarrow \bar{z}_i, c_R\leftrightarrow c_{R'}$, the contributions of $A_3$ and $B_3$ must be related. We immediately
see that
\begin{equation*}
A_3={1\ov \sqrt{k}} \frac{f_{a_1a_2a_3}}{\bar z_{12} \bar z_{13}\bar z_{23}}\langle\Phi^{(1)}_{i,i'}(x_1,\bar{x}_1)J_1^{a_1}(z_1) J_1^{a_2}(z_2) J_1^{a_3}(z_3) \Phi^{(2)}_{j,j'}(x_2,\bar{x}_2)\rangle\ .
\end{equation*}
Subsequently, we apply the Ward identity for the current $J_1^a$. When contracted with
another internal current via the Abelian term the result is zero due to the overall factor
$f_{a_1a_2a_3}$. When the contraction is via the non-Abelian part then there is no  contribution to order $1/k$.
When  $J^{a_1}$ is contracted with an external $\Phi$ one saturates the order $1/k$.
Further contraction gives rise either to pieces corresponding to disconnected diagrams or to subleading terms. Similar considerations for $B_3$ also apply.
Hence, all contributions of $\mathcal{O}(\l^3)$ are zero and we get that
\begin{equation}
\label{FF3}
\langle\Phi^{(1)}_{i,i'}(x_1,\bar{x}_1)\Phi^{(2)}_{j,j'}(x_2,\bar{x}_2)\rangle^{(3)}_{\l}=0 \ .
\end{equation}

\no
\textbf{Four-loop} $\mathcal{O}(\l^4)$: To this order the result has several terms. However, after excluding, according to our general discussion, mixed in $\l_1$ and $\l_2$ terms, we have that
  \begin{equation}
\label{order4}
 \begin{split}
  \langle\Phi^{(1)}_{i,i'}(x_1,\bar{x}_1)\Phi^{(2)}_{j,j'}(x_2,\bar{x}_2)\rangle^{(4)}_{\l}=
\frac{1}{4!\pi^4}
\int \text{d}^2z_{1234} \left(\l_1^4\, A_4 + \l_2^4\, B_4\right) ,
 \end{split}
 \end{equation}
where
\begin{equation}
A_4=\langle\Phi^{(1)}_{i,i'}J_1^{a_1}(z_1)\bar{J}^{a_1}_2(\bar{z}_1)J_1^{a_2}(z_2)\bar{J}_2^{a_2}(\bar{z}_2)J_1^{a_3}(z_3)\bar{J}_2^{a_3}(\bar{z}_3)J_1^{a_4}(z_4)\bar{J}_2^{a_4}(\bar{z}_4)\Phi^{(2)}_{j,j'}\rangle
\end{equation}
and $B_4$ follows by the parity transformation in $A_4$. To compute
$A_4$ term, we use the Ward identity for
$J_1^{a_1}$ obtaining that
\be
\label{B1}
\begin{split}
A_4&=-\frac{1}{\sqrt{k}}\frac{(t_{a_1}^{(1)})_i{}^l}{z_1-x_1}\langle\Phi^{(1)}_{l,i'}\bar{J}^{a_1}_2(\bar{z}_1)J_1^{a_2}(z_2)\bar{J}_2^{a_2}(\bar{z}_2)J_1^{a_3}(z_3)\bar{J}_2^{a_3}(\bar{z}_3)J_1^{a_4}(z_4)\bar{J}_2^{a_4}(\bar{z}_4)\Phi^{(2)}_{j,j'}\rangle
\\
&\phantom{00}-\frac{1}{\sqrt{k}}\frac{(t_{a_1}^{(2)})_j{}^l}{z_1-x_2}\langle\Phi^{(1)}_{i,i'}\bar{J}^{a_1}_2(\bar{z}_1)J_1^{a_2}(z_2)\bar{J}_2^{a_2}(\bar{z}_2)J_1^{a_3}(z_3)\bar{J}_2^{a_3}(\bar{z}_3)J_1^{a_4}(z_4)\bar{J}_2^{a_4}(\bar{z}_4)\Phi^{(2)}_{l,j'}\rangle
\\
&\phantom{00}+\frac{\delta_{a_1a_2}}{z_{12}^2}\langle\Phi^{(1)}_{i,i'}\bar{J}^{a_1}_2(\bar{z}_1)\bar{J}_2^{a_2}(\bar{z}_2)J_1^{a_3}(z_3)\bar{J}_2^{a_3}(\bar{z}_3)J_1^{a_4}(z_4)\bar{J}_2^{a_4}(\bar{z}_4)\Phi^{(2)}_{j,j'}\rangle
\\
&\phantom{0 0}+\frac{1}{\sqrt{k}}\frac{f_{a_1a_2c}}{z_{12}}\langle\Phi^{(1)}_{i,i'}\bar{J}^{a_1}_2(\bar{z}_1)J_1^{c}(z_2)\bar{J}_2^{a_2}(\bar{z}_2)J_1^{a_3}(z_3)\bar{J}_2^{a_3}(\bar{z}_3)J_1^{a_4}(z_4)\bar{J}_2^{a_4}(\bar{z}_4)\Phi^{(2)}_{j,j'}\rangle
\\
&\phantom{00}+\frac{\delta_{a_1a_3}}{z_{13}^2}\langle\Phi^{(1)}_{i,i'}\bar{J}^{a_1}_2(\bar{z}_1)J_1^{a_2}(z_2)\bar{J}_2^{a_2}(\bar{z}_2)\bar{J}_2^{a_3}(\bar{z}_3)J_1^{a_4}(z_4)\bar{J}_2^{a_4}(\bar{z}_4)\Phi^{(2)}_{j,j'}\rangle
\\
&\phantom{00}+\frac{1}{\sqrt{k}}\frac{f_{a_1a_3c}}{z_{13}}\langle\Phi^{(1)}_{i,i'}\bar{J}^{a_1}_2(\bar{z}_1)J_1^{a_2}(z_2)\bar{J}_2^{a_2}(\bar{z}_2)J_1^c(z_3)\bar{J}_2^{a_3}(\bar{z}_3)J_1^{a_4}(z_4)\bar{J}_2^{a_4}(\bar{z}_4)\Phi^{(2)}_{j,j'}\rangle
\\
&\phantom{00}+\frac{\delta_{a_1a_4}}{z_{14}^2}\langle\Phi^{(1)}_{i,i'}\bar{J}^{a_1}_2(\bar{z}_1)J_1^{a_2}(z_2)\bar{J}_2^{a_2}(\bar{z}_2)J_1^{a_3}(z_3)\bar{J}_2^{a_3}(\bar{z}_3)\bar{J}_2^{a_4}(\bar{z}_4)\Phi^{(2)}_{j,j'}\rangle
\\
&\phantom{00}+\frac{1}{\sqrt{k}}\frac{f_{a_1a_4c}}{z_{14}}\langle\Phi^{(1)}_{i,i'}\bar{J}^{a_1}_2(\bar{z}_1)J_1^{a_2}(z_2)\bar{J}_2^{a_2}(\bar{z}_2)J_1^{a_3}(z_3)\bar{J}_2^{a_3}(\bar{z}_3)J_1^c(z_4)\bar{J}_2^{a_4}(\bar{z}_4)\Phi^{(2)}_{j,j'}\rangle \ .
\end{split}
\ee
The first two lines of \eqn{B1} correspond to a contraction of $J_1^{a_1}$ with the external
fields and the rest with its contraction, Abelian and non-Abelian,
with $J^{a_2}$, $J^{a_3}$ and $J^{a_4}$.

The first line of \eqn{B1} after further contraction with $J_1^{a_2}$
and subsequently with $J_1^{a_3}$ and $J_1^{a_4}$
gives up to order $1/k$ the expression
    \be
    \begin{split}
&\frac{1}{k}\left(\frac{(t_{a_1})_i{}^l(t_{a_2})_l{}^m}{(z_1-x_1)(z_2-x_1)z_{34}^2}\langle\Phi^{(1)}_{m,i'}\bar{J}_2^{a_1}(\bar{z}_1) \bar{J}_2^{a_2}(\bar{z}_2) \bar{J}_2^{a_3}(\bar{z}_3) \bar{J}_2^{a_3}(\bar{z}_4)  \Phi^{(2)}_{j,j'}\rangle\right.\\
&-\frac{(t_{a_1})_i{}^l(t_{a_2}^*)_j{}^m}{(z_1-x_1)(z_2-x_2)z_{34}^2}\langle\Phi^{(1)}_{l,i'}\bar{J}_2^{a_1}(\bar{z}_1) \bar{J}_2^{a_2}(\bar{z}_2) \bar{J}_2^{a_3}(\bar{z}_3) \bar{J}_2^{a_3}(\bar{z}_4)  \Phi^{(2)}_{m,j'}\rangle\\
&+\frac{(t_{a_1})_i{}^l(t_{a_4})_l{}^m}{(z_1-x_1)(z_4-x_1)z_{23}^2}\langle\Phi^{(1)}_{m,i'}\bar{J}_2^{a_1}(\bar{z}_1) \bar{J}_2^{a_2}(\bar{z}_2) \bar{J}_2^{a_2}(\bar{z}_3) \bar{J}_2^{a_4}(\bar{z}_4)  \Phi^{(2)}_{j,j'}\rangle\\
&-\frac{(t_{a_1})_i{}^l(t_{a_4}^*)_j{}^m}{(z_1-x_1)(z_4-x_2)z_{23}^2}\langle\Phi^{(1)}_{l,i'}\bar{J}_2^{a_1}(\bar{z}_1) \bar{J}_2^{a_2}(\bar{z}_2) \bar{J}_2^{a_2}(\bar{z}_3) \bar{J}_2^{a_4}(\bar{z}_4)  \Phi^{(2)}_{m,j'}\rangle\\
&-\frac{(t_{a_1})_i{}^l f_{a_2a_3a_4}}{(z_1-x_1)z_{23}z_{34}^2}\langle\Phi^{(1)}_{l,i'}\bar{J}_2^{a_1}(\bar{z}_1) \bar{J}_2^{a_2}(\bar{z}_2) \bar{J}_2^{a_3}(\bar{z}_3) \bar{J}_2^{a_4}(\bar{z}_4)  \Phi^{(2)}_{j,j'}\rangle\\
&+\frac{(t_{a_1})_i{}^l(t_{a_3})_l{}^m}{(z_1-x_1)(z_3-x_1)z_{24}^2}\langle\Phi^{(1)}_{m,i'}\bar{J}_2^{a_1}(\bar{z}_1) \bar{J}_2^{a_2}(\bar{z}_2) \bar{J}_2^{a_3}(\bar{z}_3) \bar{J}_2^{a_2}(\bar{z}_4)  \Phi^{(2)}_{j,j'}\rangle\\
&-\frac{(t_{a_1})_i{}^l(t_{a_3}^*)_j{}^m}{(z_1-x_1)(z_3-x_2)z_{24}^2}\langle\Phi^{(1)}_{l,i'}\bar{J}_2^{a_1}(\bar{z}_1) \bar{J}_2^{a_2}(\bar{z}_2) \bar{J}_2^{a_3}(\bar{z}_3) \bar{J}_2^{a_2}(\bar{z}_4)  \Phi^{(2)}_{m,j'}\rangle\\
&\left.-\frac{(t_{a_1})_i{}^lf_{a_2a_4a_3}}{(z_1-x_1)z_{24}z_{34}^2}\langle\Phi^{(1)}_{l,i'}
\bar{J}_2^{a_1}(\bar{z}_1) \bar{J}_2^{a_2}(\bar{z}_2) \bar{J}_2^{a_3}(\bar{z}_3)
\bar{J}_2^{a_4}(\bar{z}_4)  \Phi^{(2)}_{j,j'}\rangle    \right)\ ,
\label{firstline}
\end{split}
\ee
where we have only kept terms corresponding to connected diagrams.
From the second line of (\ref{B1}), a similar expression occurs.
The contribution from the third line of \eqn{B1}, which results from an Abelian contraction among currents, is

\begin{align}
\begin{split}
&\frac{1}{k}\left(\frac{(t_{a_3})_i{}^l(t_{a_4})_l{}^m}{(z_3-x_1)(z_4-x_1)z_{12}^2} \langle\Phi^{(1)}_{m,i'}\bar{J}_2^{a_1}(\bar{z}_1) \bar{J}_2^{a_1}(\bar{z}_2) \bar{J}_2^{a_3}(\bar{z}_3) \bar{J}_2^{a_4}(\bar{z}_4)  \Phi^{(2)}_{j,j'}\rangle\right.\\
&-\frac{(t_{a_3})_i{}^l(t_{a_4}^*)_j{}^m}{(z_3-x_1)(z_4-x_2)z_{12}^2} \langle\Phi^{(1)}_{l,i'}\bar{J}_2^{a_1}(\bar{z}_1) \bar{J}_2^{a_1}(\bar{z}_2) \bar{J}_2^{a_3}(\bar{z}_3) \bar{J}_2^{a_4}(\bar{z}_4)  \Phi^{(2)}_{m,j'}\rangle\\
&-\frac{(t_{a_3}^*)_j{}^l(t_{a_4})_i{}^m}{(z_3-x_2)(z_4-x_1)z_{12}^2}\langle\Phi^{(1)}_{m,i'}\bar{J}_2^{a_1}(\bar{z}_1) \bar{J}_2^{a_1}(\bar{z}_2) \bar{J}_2^{a_3}(\bar{z}_3) \bar{J}_2^{a_4}(\bar{z}_4)  \Phi^{(2)}_{l,j'}\rangle\\
&+\frac{(t_{a_3}^*)_j{}^l(t_{a_4}^*)_l{}^m}{(z_3-x_2)(z_4-x_2)z_{12}^2}\langle\Phi^{(1)}_{i,i'}\bar{J}_2^{a_1}(\bar{z}_1) \bar{J}_2^{a_1}(\bar{z}_2) \bar{J}_2^{a_3}(\bar{z}_3) \bar{J}_2^{a_4}(\bar{z}_4)  \Phi^{(2)}_{m,j'}\rangle\\
&-\frac{f_{a_3a_4c}(t_{a_c})_i{}^l}{(z_4-x_1)z_{34}z_{12}^2}\langle\Phi^{(1)}_{l,i'}\bar{J}_2^{a_1}(\bar{z}_1) \bar{J}_2^{a_1}(\bar{z}_2) \bar{J}_2^{a_3}(\bar{z}_3) \bar{J}_2^{a_4}(\bar{z}_4)  \Phi^{(2)}_{j,j'}\rangle\\
&\left.+\frac{f_{a_3a_4c}(t_{a_c}^*)_j{}^l}{(z_4-x_2)z_{34}z_{12}^2}\langle\Phi^{(1)}_{i,i'}
\bar{J}_2^{a_1}(\bar{z}_1) \bar{J}_2^{a_1}(\bar{z}_2) \bar{J}_2^{a_3}(\bar{z}_3)
\bar{J}_2^{a_4}(\bar{z}_4)  \Phi^{(2)}_{l,j'}\rangle \right)
\ .\label{thirdline}
\end{split}
\end{align}
The rest of the terms arising from Abelian contractions among currents in \eqn{B1}
give similar results.

\no
Finally, the fourth line of \eqn{B1}, resulting from  non-Abelian contraction among currents gives
\begin{align}
\begin{split}
&-\frac{1}{k}\left(\frac{(t_{a_c})_i{}^lf_{a_1a_2c}}{(z_2-x_1)z_{12}}\frac{\delta_{a_3a_4}}{z_{34}^2}\langle\Phi^{(1)}_{l,i'}\bar{J}_2^{a_1}(\bar{z}_1) \bar{J}_2^{a_2}(\bar{z}_2) \bar{J}_2^{a_3}(\bar{z}_3) \bar{J}_2^{a_4}(\bar{z}_4)  \Phi^{(2)}_{j,j'}\rangle\right.\\
&-\frac{(t_{a_c}^*)_j{}^lf_{a_1a_2c}}{(z_2-x_2)z_{12}}\frac{\delta_{a_3a_4}}{z_{34}^2}\langle\Phi^{(1)}_{i,i'}\bar{J}_2^{a_1}(\bar{z}_1) \bar{J}_2^{a_2}(\bar{z}_2) \bar{J}_2^{a_3}(\bar{z}_3) \bar{J}_2^{a_4}(\bar{z}_4)  \Phi^{(2)}_{l,j'}\rangle\\
&+\frac{(t_{a_4})_i{}^lf_{a_1a_2a_3}}{(z_4-x_1)z_{12}z_{23}^2}\langle\Phi^{(1)}_{l,i'}\bar{J}_2^{a_1}(\bar{z}_1) \bar{J}_2^{a_2}(\bar{z}_2) \bar{J}_2^{a_3}(\bar{z}_3) \bar{J}_2^{a_4}(\bar{z}_4)  \Phi^{(2)}_{j,j'}\rangle\\
&-\frac{(t_{a_4}^*)_j{}^lf_{a_1a_2a_3}}{(z_4-x_2)z_{12}z_{23}^2}\langle\Phi^{(1)}_{i,i'}\bar{J}_2^{a_1}(\bar{z}_1) \bar{J}_2^{a_2}(\bar{z}_2) \bar{J}_2^{a_3}(\bar{z}_3) \bar{J}_2^{a_4}(\bar{z}_4)  \Phi^{(2)}_{l,j'}\rangle\\
&+\frac{(t_{a_3})_i{}^lf_{a_1a_2a_4}}{(z_3-x_1)z_{12}z_{24}^2}\langle\Phi^{(1)}_{l,i'}\bar{J}_2^{a_1}(\bar{z}_1) \bar{J}_2^{a_2}(\bar{z}_2) \bar{J}_2^{a_3}(\bar{z}_3) \bar{J}_2^{a_4}(\bar{z}_4)  \Phi^{(2)}_{j,j'}\rangle\\
&\left.-\frac{(t_{a_3}^*)_j{}^lf_{a_1a_2a_4}}{(z_3-x_2)z_{12}z_{24}^2}\langle\Phi^{(1)}_{i,i'}
\bar{J}_2^{a_1}(\bar{z}_1) \bar{J}_2^{a_2}(\bar{z}_2) \bar{J}_2^{a_3}(\bar{z}_3) \bar{J}_2^{a_4}(\bar{z}_4)  \Phi^{(2)}_{l,j'}\rangle \right)\ ,
 \label{fourthline}
\end{split}
\end{align}
where we have once again kept only terms corresponding to connected diagrams.
The rest of the terms resulting from non-Abelian contraction among currents in \eqn{B1} give
similar results.

\no
In the above expressions a 6-point function appears with two external fields and four currents.
Since we have already saturated the factor $1/k$, in order to evaluate it we can simply contract the currents among themselves
via the Abelian part of their OPE. The result, to ${\cal O}(1)$ in which we are interested,  is
\begin{equation}
\begin{split}
&
\langle\Phi^{(1)}_{i,i'}\bar{J}_2^{a_1}(\bar{z}_1)
 \bar{J}_2^{a_2}(\bar{z}_2) \bar{J}_2^{a_3}(\bar{z}_3) \bar{J}_2^{a_4}(\bar{z}_4)
 \Phi^{(2)}_{j,j'}\rangle
=\bigg( \frac{\delta_{a_1a_2}\delta_{a_3a_4}}{\bar{z}_{12}^2
\bar{z}_{34}^2}+ \frac{\delta_{a_1a_3}\delta_{a_2a_4}}{\bar{z}_{13}^2 \bar{z}_{24}^2}
+ \frac{\delta_{a_1a_4}\delta_{a_2a_3}}{\bar{z}_{14}^2 \bar{z}_{23}^2}\bigg)
\\
&\qq\qq\qq\qq\qq\qq\qq\qq\qq
\times \frac{(\mathbb{I}_R\otimes \mathbb{I}_{R'})_{ii',jj'}}{x_{12}^{2\D_R}
\bar{x}_{12}^{2\bar{\D}_{R'}}}\ .
\end{split}
\end{equation}
Using the above in \eqn{order4}, by taking into account the symmetry under
the exchange of external points $x_1\leftrightarrow x_2$ and contracting the group indices we get
that
\begin{equation}
\label{FF4}
\langle\Phi^{(1)}_{i,i'}(x_1,\bar{x}_1)\Phi^{(2)}_{j,j'}
(x_2,\bar{x}_2)\rangle^{(4)}_{\l}={1\ov k}\big(c_R\l_1^4+ c_{R'}\l_2^4\big)
\frac{(\mathbb{I}_R\otimes \mathbb{I}_{R'})_{ii',jj'}}{x_{12}^{2\D_R}
\bar{x}_{12}^{2\bar{\D}_{R'}}}
\ln\frac{\varepsilon^2}{|x_{12}|^2}
\ .
\end{equation}
Adding together the perturbative results of \eqref{FF1}, \eqref{FF2}, \eqref{FF3} and \eqref{FF4},  we find 
the two-point function for primaries up to order $\mathcal{O}(\l^4/k)$ to be
\be
\label{fgjh}
\begin{split}
& \langle\Phi^{(1)}_{i,i'}(x_1,\bar{x}_1)\Phi^{(2)}_{j,j'}(x_2,\bar{x}_2)\rangle_\l
={1\ov k}\big(c_R\l_1^2(1+\l_1^2)+ c_{R'}\l_2^2(1+\l_2^2)\big)
\\
&\qq\qq\qq\qq \times
\frac{(\mathbb{I}_R\otimes
\mathbb{I}_{R'})_{ii',jj'}}{x_{12}^{2\D_R}\bar{x}_{12}^{2\bar{\D}_{R'}}}
\ln\frac{\varepsilon^2}{|x_{12}|^2} + {\cal O}(\l^5/k) \ ,
\end{split}
\ee
from which we extract the perturbative expression \eqn{prim-dim-pert} in the main text.


\end{document}